\documentclass[twocolumn,amsmath,trackchanges]{aastex702}

\usepackage{ulem}
\usepackage{bm}

\shorttitle{Leptonic and hadronic $\gamma$-rays from coronae}
\shortauthors{Le Bihan, Gro{\v{s}}elj \& Lemoine}

\submitjournal{ApJL}

\begin{document}
\title{Gamma-rays from black hole coronae: disentangling the leptonic and hadronic contributions}

\author[orcid=0009-0008-9453-6266]{S\'ebastien Le Bihan}
\affiliation{Astroparticule et Cosmologie, CNRS, Universit\'e Paris Cit\'e, F-75013 Paris, France}
\email[show]{lebihan@apc.in2p3.fr}

\author[orcid=0000-0002-5408-3046]{Daniel Gro{\v{s}}elj}
\affiliation{Centre for mathematical Plasma Astrophysics, Department of Mathematics, KU Leuven, B-3001 Leuven, Belgium}
\email{daniel.groselj@kuleuven.be}

\author[orcid=0000-0002-2395-7812]{Martin Lemoine}
\affiliation{Astroparticule et Cosmologie, CNRS, Universit\'e Paris Cit\'e, F-75013 Paris, France}
\email{mlemoine@apc.in2p3.fr}

\correspondingauthor{S\'ebastien Le Bihan}
\begin{abstract}
IceCube detections of TeV neutrinos from nearby Seyfert galaxies imply, if confirmed, that the accompanying gamma rays are substantially reprocessed down to the MeV range by electromagnetic cascades, which places multi-messenger constraints on neutrino emission models. However, gamma rays can also be produced by the primary nonthermal electrons of the X-ray corona, as recent radiative kinetic simulations of magnetized turbulence have shown. To disentangle the leptonic and hadronic contributions, we self-consistently combine results from these simulations with predictions for hadronically induced electromagnetic cascades. Focusing on NGC 4151, we find that the leptonic channel dominates the MeV band, and remains comparable to the hadronic cascade up to $\lesssim$0.1 GeV, when the nonthermal electron tail carries $\gtrsim5\%$ of the electron energy. The leptonic contribution may be detectable in soft gamma rays by the upcoming Compton Spectrometer and Imager (COSI) mission, whereas our predicted hadronic component falls below its sensitivity.
We also find that the same nonthermal electron population can account for both the $0.1-1$ MeV excess of the cosmic X-ray background, generally attributed to AGNs, and the MeV tail of Cygnus~X-1 in the hard state, lending observational support to our model.
Our results offer a refined view of multi-messenger signals from black-hole coronae, grounded in their nonthermal kinetic physics.
\end{abstract}

\section{Introduction} \label{sec:intro}
Accreting black holes are prominent sources of X-ray radiation across a wide range of mass scales, from stellar-mass black holes in X-ray binaries to supermassive black holes powering active galactic nuclei (AGN). The hard X-ray continuum radiation is believed to originate in a semi-relativistic, turbulent magnetized corona within a few tens of gravitational radii, where dissipative processes energize electrons (and positrons) that near-instantly radiate their energy through inverse Compton interactions
\citep[e.g.][]{1985ApJ...294L..79Z,1991ApJ...380L..51H,1996ApJ...470..249P,2011MNRAS.414.3330V,2015MNRAS.451.4375F,beloborodovRadiativeMagneticReconnection2017}.

Evidence for proton co-energization had long been missing until the IceCube report of an excess of $\sim 1$--$10$\,TeV neutrinos spatially associated with the Seyfert\,2 galaxy NGC\,1068 \citep{aartsenTimeIntegratedNeutrinoSource2020a,icecubecollaborationEvidenceNeutrinoEmission2022}, at a (post-trial) $4.2\sigma$ confidence level, together with growing evidence for emission from NGC\,4151 and NGC\,7469 \citep{abbasiseyfert2025,abbasiEvidenceNeutrinoEmission2025}, at (pre-trial) $3.1\sigma$ and $3.8\sigma$ respectively. The lack of $\gamma$-rays at a level comparable to the neutrino flux points toward a compact, $\gamma\gamma$-thick production region \citep[e.g.][]{inoueHighenergyParticlesAccretion2019,muraseHiddenCoresActive2020,2021ApJ...922...45K,eichmannSolvingMultimessengerPuzzle2022, dasRevealingProductionMechanism2024}, best identified with the turbulent, magnetized corona of the accreting black hole. Detailed theoretical investigations further indicate that the black hole corona offers the requisite conditions to power up the protons up to the implied energies ($\sim 100\,$TeV) through stochastic (turbulent) acceleration, with a sufficient luminosity to accommodate the observed neutrino flux~\citep[e.g.][]{muraseHiddenCoresActive2020,2022ApJ...941L..17M,eichmannSolvingMultimessengerPuzzle2022,mbarekInterplayAcceleratedProtons2024,lemoineNeutrinosStochasticAcceleration2025,2025ApJ...995..166Y,SLB26}. If confirmed, these observations point toward non-jetted AGN coronae as significant contributors to the extragalactic neutrino flux~\citep{padovaniHighenergyNeutrinosVicinity2024,neronovNeutrinoSignalPopulation2024,2025ApJ...989..215F,2026arXiv260220145M,2026arXiv260220969Y, Karavola2026} and open a novel hadronic window on the physics of the corona.

The high $\gamma\gamma$ opacity implies an efficient reprocessing of proton energy lost in Bethe-Heitler and hadronic interactions into the soft gamma-ray range ($\sim 1-10\,$MeV), and thereby offers a promising signature of hadronic processes in black hole coronae~\citep{muraseHiddenCoresActive2020}. However, MeV tails of leptonic origin are also predicted by hybrid corona models invoking a population of nonthermal primary electrons~\citep[e.g.][and references therein]{Ghisellini+93,2014SSRv..183...61P,2017MNRAS.467.2566F,Zdziarski2017}. Recent {\it ab initio} kinetic numerical simulations of turbulent plasmas representative of black hole coronae have further demonstrated that electron acceleration to nonthermal energies is in fact generic~\citep{groseljRadiativeParticleinCellSimulations2024,2024NatCo..15.7026N,groseljHighenergyEmissionTurbulent2026}.  

To disentangle these contributions, we leverage in this Letter recent advances in kinetic simulations and in theoretical modeling of proton acceleration, which we integrate with extensive multi-messenger data to establish robust predictions for the leptonic and hadronic contributions to the BH corona flux in the MeV -- GeV range. Our predictions are especially aimed at the upcoming Compton Spectrometer and Imager (COSI)~\citep{2024icrc.confE.745T}, which will provide high-sensitivity  measurements in the $0.2-5\,$MeV range. We focus on the Seyfert 1/1.5 NGC\,4151 that combines strong X-ray coverage thanks to its weak obscuration, substantial evidence for high-energy neutrino production, and status as a prime COSI target. In detail, we develop a theoretical model for the primary electron nonthermal population informed by recent PIC simulations \citep{groseljRadiativeParticleinCellSimulations2024,2024NatCo..15.7026N,groseljHighenergyEmissionTurbulent2026}, which we combine with the \texttt{Turb-AM3} code \citep{klingerAM$^3$OpenSourceTool2024,SLB26} to incorporate in a self-consistent manner proton acceleration and all relevant hadronic and electromagnetic interaction processes. Comparing these predictions to existing observational constraints allows us to gauge the different contributions to the spectral shape in the COSI range.

We conclude that the radiative output of nonthermal primary electrons substantially outweighs that of secondary electromagnetic cascades of hadronic origin, up to sub-GeV energies, and show that COSI has the potential to probe this nonthermal population of electrons. Cumulating these predicted emission spectra over the AGN population, we further demonstrate that this same model offers a satisfactory account of the hard cosmic X-ray background, as anticipated early on \citep{2008ApJ...672L...5I}, which lends further support to the picture developed here.

\section{Nonthermal populations} \subsection{Primary electrons and positrons}\label{sec:electrons}
Particle-in-cell simulations of turbulent, magnetized electron-ion coronae incorporating inverse Compton and synchrotron radiative channels offer a first-principles characterization of the steady-state electron energy distribution under conditions directly relevant to AGN coronae, meaning a proton magnetization parameter $\sigma_p \simeq 0.1$, where $\sigma_p\equiv u_B/(n_p m_pc^2)\simeq (v_{\rm A}/c)^2$ corresponds to the ratio of magnetic to plasma energy density, and $v_{\rm A}$ to the Alfvénic velocity; the plasma is also assumed to be strongly turbulent, meaning a fluctuation amplitude $\delta B/B_0\sim O(1)$ on the outer scale. These simulations reveal a quasi-thermal (Maxwellian) component with an effective temperature $\Theta_e\equiv k_BT_e/m_e c^2 \sim O(0.2)$, together with a high-energy nonthermal tail, which is well described by a power law with index $p \approx 3.5$~\citep{groseljHighenergyEmissionTurbulent2026}. In these environments of high compactness $\ell \equiv L\sigma_{\rm T}/(R_{\rm c} m_ec^3)$~\citep{1983MNRAS.205..593G}, with $L$ and $R_{\rm c}$ the luminosity and the radius of the corona, electrons radiate their energy on a timescale shorter than an eddy turn-around time. In these fast cooling conditions, electron energization predominantly takes place in intense, intermittent reconnecting current sheets embedded in the turbulent flow, rather than through stochastic acceleration.
This is corroborated by the measured pitch-angle distribution, which shows nonthermal electrons preferentially aligned with the local magnetic field, as also found in kinetic simulations of reconnection at comparable magnetization. These simulations, although they do not include radiative losses, yield a similar power-law index $p\approx3.0-3.5$~\citep{comissoConcurrentParticleAcceleration2024}.

\begin{figure}
\includegraphics[width=\columnwidth]{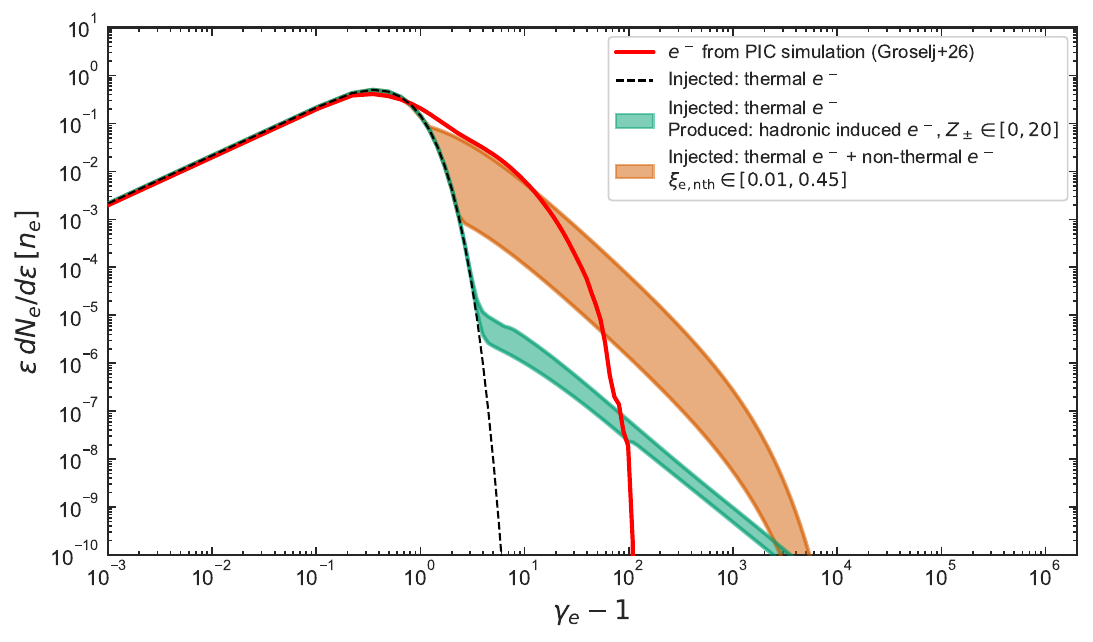}
\caption{Electron kinetic energy [$\varepsilon\equiv(\gamma_e-1)m_ec^2$] distributions considered in this work: in dashed black, a purely thermal distribution for primary electrons; in green, a thermal distribution and the secondary electrons of hadronic origin, discussed in Sec.~\ref{sec:protons}; in orange, a thermal distribution continuously connected to a nonthermal $p=3$ power law with $\xi_{e,\rm nth} \in [0.01, 0.45]$; in red, the spectrum directly extracted from the radiative PIC simulations of \citet{groseljHighenergyEmissionTurbulent2026}. 
\label{fig:electron}}
\end{figure}

The nonthermal electrons observed in kinetic simulations of turbulent coronal plasmas carry an energy density comparable to that of the thermal bath, making them energetically significant rather than a minor spectral correction. This nonthermal population, shown by the thick red line in Fig.~\ref{fig:electron}, directly shapes the predicted emission above the $\sim 100$\,keV Comptonization peak, producing an MeV tail that hardens with the electron magnetization $\sigma_e\equiv u_B/(n_em_ec^2)$. Despite being the largest to date, this kinetic simulation is subject to limitations imposed by its high numerical cost: (1) it assumed an electron-proton mixture of equal densities ($n_e=n_p$) and left aside pair production and annihilation channels; (2) it used a proton-to-electron mass ratio of $m_p/m_e=144$; (3) it is 2D3V, meaning two spatial dimensions, three in velocity space. With regard to (1), we retain in the following the pair loading factor $Z_\pm = n_e/n_p - 1$ as a free parameter, noting that $n_e$ ($n_p$) here represents the total electron and positron (respectively, proton) number density. We anticipate $Z_\pm \sim O(1)$ but discuss possible consequences where relevant. Note that $\sigma_p = (m_e/m_p)(1+Z_\pm)\sigma_e$ and $\sigma_p\simeq 0.1$ imply $\sigma_e\simeq 180/(1+Z_\pm)$. The restricted mass ratio (2) implies a reduced $\sigma_e$ at a given $\sigma_p$, and correspondingly a reduced compactness, since $\ell\approx 2\tau_{\rm T}\sigma_e (1-q_p)(v_{\rm A}/c)$  in terms of the optical depth $\tau_{\rm T}\sim O(1)$ and the fraction $q_p$ of turbulent power going into proton energization~\citep{groseljHighenergyEmissionTurbulent2026}. For reference the simulation spectrum shown in Fig.~\ref{fig:electron} (red line) corresponds to $\sigma_p=0.19$ and $\ell= 8.5$. The reduced dimensionality (3) likely underestimates the rate of energization by reconnection~\citep{2023ApJ...956L..36Z}. The electron nonthermal tail observed by \citet{groseljHighenergyEmissionTurbulent2026} displays a sharp cut-off around a Lorentz factor $\gamma_{e,\rm cut}\sim 30$, and we regard it as a strict lower limit.

To account for these various effects, we extend our analysis to nonthermal powerlaw electron spectra with a larger maximal energy and a smaller injection fraction. The cut-off $\gamma_{e,\rm cut}$ is bounded from above by radiative losses, inverse Compton and synchrotron alike, as well as by a confinement limit due to the finite size of the accelerating current sheets. Using Eq.(11) of  \citet{groseljHighenergyEmissionTurbulent2026} for the limit imposed by radiative losses, one finds $\gamma_{e,\rm cut}\sim O(10^5-10^6)$ for typical AGN coronae. However, the confinement limit is likely more constraining. Statistical studies of the length ($l_{\rm CS}$), width ($w_{\rm CS}$) and thickness of current sheets in a turbulent plasma reveal a continuous spectrum of length scales with aspect ratios of the order of $l_{\rm CS}/d_{\rm p}\gtrsim 100-300$ and $w_{\rm CS}/d_{\rm p}\gtrsim 10-50$ extending up to the largest scale in the system~\citep[e.g.][]{2015PhPl...22d2902M}; they are written here in units of $d_{\rm p}$, the ion skin depth. Assuming that acceleration proceeds at a rate $\nu_{\rm acc}\simeq \beta_{\rm rec}(v_{\rm A}/c) \omega_{\rm g}$, where $\omega_{\rm g}=e B/(\gamma_e m_e c)$ is the gyrofrequency~\citep{2023ApJ...956L..36Z}, the confinement limit reads $\gamma_{e,\rm cut}\simeq \langle w_{\rm CS}/d_e\rangle (v_{\rm A}/c)\beta_{\rm rec}\sigma_e^{1/2}$. For $v_{\rm A}/c\simeq 0.3$ ($\sigma_p\simeq 0.1$), $\sigma_e \simeq (1+Z_\pm)^{-1} (m_p/m_e)\sigma_p$ and $d_p/d_e = (1+Z_\pm)^{1/2}(m_p/m_e)^{1/2}$, this gives $\gamma_{e,\rm cut}\simeq 10^3 \,\langle w_{\rm CS,min}/d_p\rangle/50 $, independent of $Z_\pm$. We adopt $\gamma_{e,\rm cut}\simeq 10^3$.

The nonthermal fraction is mostly governed, in the absence of cooling, by the fraction of plasma that current sheets process in an eddy turn-around time~\citep[e.g.][]{comissoInterplayMagneticallyDominated2019a}. Consequently, in the fast cooling regime, one expects this fraction to decrease by the ratio of the turn-around time to the cooling time for thermal electrons, and hence in proportion to the compactness $\ell$. In BH coronae of compactness $\ell\gtrsim 10$, the nonthermal electron fraction may thus be lower than that seen in the kinetic simulations that probe $\ell < 10$.

In this Letter, we thus consider three sets of electron distributions, shown in Fig.~\ref{fig:electron}. The first (red) is the spectrum extracted directly from the radiative PIC simulations of \citet{groseljHighenergyEmissionTurbulent2026}: a thermal population plus a nonthermal tail extending up to $\gamma_{e,\rm cut}\sim 30$. The second (orange) models a Maxwell--J\"uttner distribution at temperature $\Theta_e$ continuously connected, at a Lorentz factor $\gamma_{e,\rm nth}$, to a nonthermal power law of index $p = 3.0$ with an exponential cutoff at $\gamma_{e,\rm cut} = 10^3$, as defined by \citet{2009MNRAS.400..330G}. The junction $\gamma_{e,\rm nth}$ is set by the nonthermal fraction $\xi_{e,\rm nth}$, defined as the ratio of the kinetic energy density of electrons with $\gamma_e > \gamma_{e,\rm nth}$ to the total electron kinetic energy density; for $\Theta_e\simeq 0.2$, the range $\xi_{e,\rm nth} = 0.01-0.45$ spanned by the orange band corresponds to $\gamma_{e,\rm nth}\simeq 3.4-2.2$. 
With the same definition, the PIC spectrum corresponds to $\xi_{e,\rm nth}\simeq 0.43$.
The third (green) assumes purely thermal primary electrons; its nonthermal extension is the secondary electron (and positron) population produced by hadronic processes (Sec.~\ref{sec:protons}), and the band width reflects the uncertainty on the hadronic component.

\subsection{Protons and secondaries}\label{sec:protons}
We use the \texttt{Turb-AM3} numerical framework, presented in detail in \citet{SLB26}, to self-consistently track the co-evolution of protons, electrons (+positrons), photons, neutrinos, pions and muons produced by hadronic interactions, Bethe-Heitler and Breit-Wheeler pair production, pair annihilation, synchrotron, inverse Compton and decay channels in a one-zone, turbulent magnetized AGN corona of size $R_{\rm c}$, threaded by a magnetic field $\delta B$ and fed by an accretion disk of luminosity $L_d$. Protons are stochastically accelerated by turbulent electric fields, as described either by a diffusive Fokker-Planck model or by a generalized Fermi description of stochastic particle acceleration in large-amplitude turbulence~\citep{lemoineFirstPrinciplesFermiAcceleration2022,2025PhRvE.112a5205L}. 

The nonthermal proton content is parametrized by the injection fraction $\xi_{\rm p,\rm inj} = P_{\rm p,nth}/P_{\rm p,tot}$, expressed as the ratio of the pressure of protons injected into the acceleration process relative to that in the thermal proton bath.
For each  species, the code solves a time-dependent transport equation in energy space describing stochastic acceleration (for protons), together with the relevant radiative and hadronic interactions, and escape channels.
It evolves up to a maximal time $t_{\rm adv}$ characterizing advection out of the corona, either through inflow into the black hole or with an outflow. Turbulent dissipation into particle acceleration is coupled back to a cascade equation for the magnetic power spectrum, so that the acceleration efficiency self-consistently decreases as energy is extracted into nonthermal protons \citep{lemoineNonlinearAspectsStochastic2024,lemoineNeutrinosStochasticAcceleration2025}. This effectively bounds the power carried by nonthermal protons, subsequent to acceleration, to a fraction $q_{p, \rm nth} \lesssim 1$ of the available turbulent power $u_B/(l_{\rm coh}/v_{\rm A})$, where $l_{\rm coh}$ denotes the outer scale of the turbulence, and $l_{\rm coh}/v_{\rm A}$ the characteristic turn-around time on that scale. 

The primary electron (and positron) content controls the hard X-ray radiative background that controls the rate of $p\gamma$ interactions, as well as the background proton density (at given $\tau$ and $Z_\pm$), which sets the rate of hadronic $pp$ interactions. We fix here the electron distribution according to the prescriptions of Sec.~\ref{sec:electrons}. This primary electron distribution is not evolved in time, because it is assumed to reach a steady state on timescales $\ll R_{\rm c}/c$, as a result of the high compactness. This contrasts with protons, whose distribution evolves on timescales $\gg R_{\rm c}/c$. For each electron spectral model, the electron density $n_e$ is set so as to keep the Compton parameter $y_\mathrm{C} = \sigma_T R_{\rm c}\int \gamma_e^2\,\mathrm{d}n_e$ fixed, ensuring our prescriptions yield approximately the same hard X-ray spectral shape. To obtain a finer description of the Comptonization process in the sub-relativistic range of electron energies, we have modified the original (relativistic) inverse Compton kernel of \texttt{AM3} to incorporate the energy independent kernel description from \citet{VP09}\footnote{In the present case of a coronal environment this modification has only a minor role. The precise study of the implication of a relativistic vs trans-relativistic description of AM3 will form the base of a future work.}. 

Overall, the free parameters of the model are the coronal radius $R_{\rm c}$, the turbulent coherence length $l_{\rm coh}$, the Alfv\'en velocity $v_\mathrm{A}$, the nonthermal proton injection fraction $\xi_{p,\rm inj}$, and the pair loading factor $Z_\pm$. These combine into the characteristic timescales of advection, diffusive escape, and proton acceleration that essentially govern the high-energy proton spectral shape, and hence that of hadronic electromagnetic cascades. Note that $\xi_{p,\rm inj}$ is not a tightly constrained parameter: for $\xi_{p,\rm inj} \gtrsim 0.03$, the self-regulation of acceleration by turbulent damping decides the overall normalization of the nonthermal proton distribution, independently of the initial value of $\xi_{p,\rm inj}$; the aforementioned radiative PIC simulations indeed point to such a ratio being close to unity.
Finally, the code self-consistently determines the amount of secondary pairs produced by hadronic electromagnetic cascades. Over a broad range of parameters, we find that the corresponding pair loading factor is on average well below unity, and in more extreme cases of order unity, thus within the range of values of $Z_\pm$ that we explore in the following.

\begin{figure*}
\resizebox{\hsize}{!}{\includegraphics[clip=true]{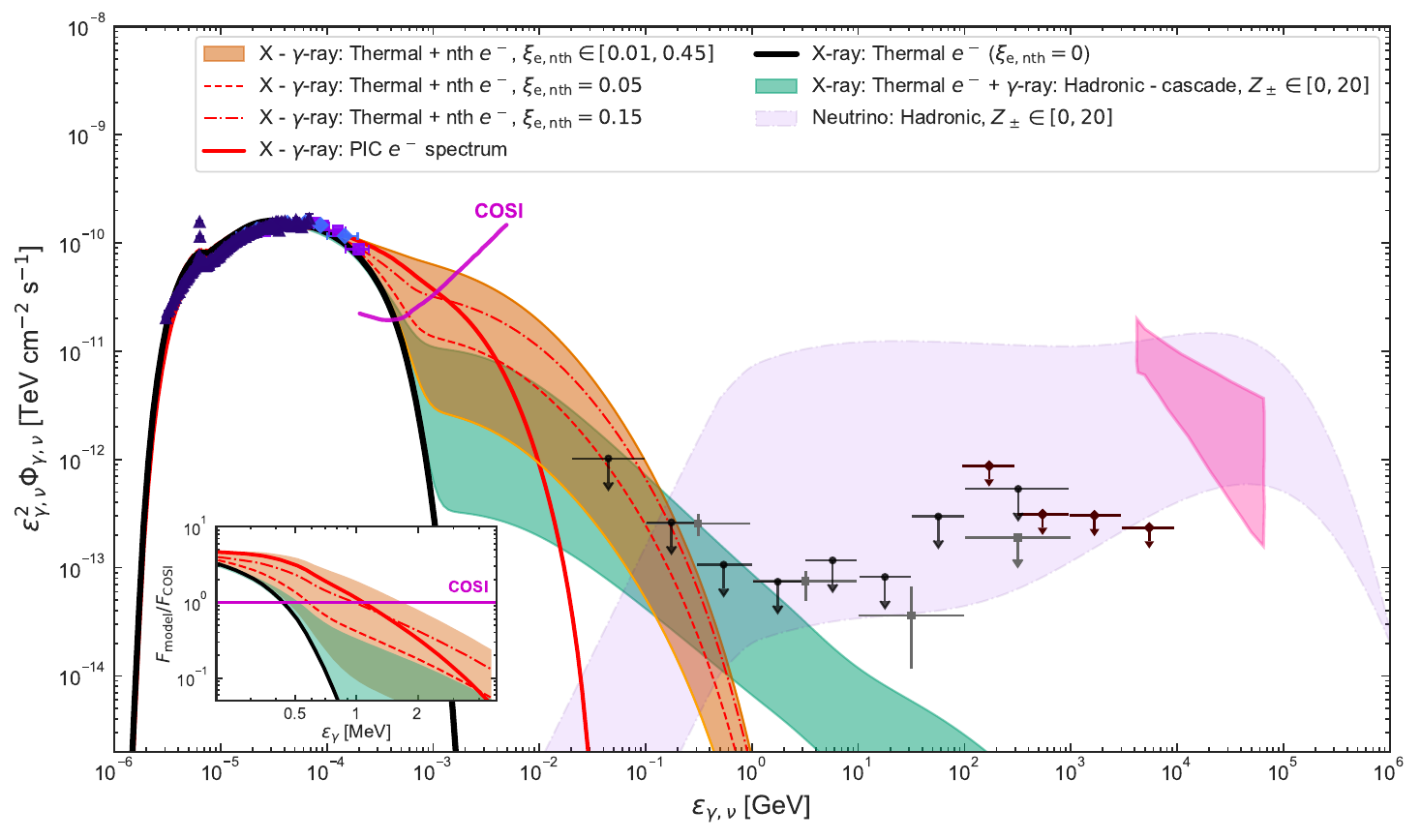}}
\caption{Multi-messenger SED of NGC~4151 for different electron models.
\textit{Leptonic only} (hadronic processes switched off): the orange band shows the photon output for thermal + nonthermal electrons with a nonthermal energy fraction $\xi_{\rm e,  nth} \in [0.01, 0.45]$, with $\xi_{\rm e, nth} = 0.05$ and $0.15$ highlighted as red dashed and dash-dotted lines; the red solid line shows the output for the PIC electron distribution.
\textit{Thermal electrons plus hadronic processes}: the green (photons) and purple (neutrinos) bands span pair-enrichment fractions $Z_\pm \in [0, 20]$ (the high ends of the bands correspond to $Z_\pm = 0$).
The inset plot shows the fluxes of the different models, normalized to the COSI sensitivity.
\textit{Thermal electrons only}: black line.
Data: archival X-ray spectra from \textit{INTEGRAL} (purple squares), \textit{Swift}-BAT (blue diamonds), and \textit{NuSTAR} (dark blue triangles), normalized to \textit{Swift}-BAT; \textit{Fermi}-LAT upper limits (black arrows; \citealt{muraseSubGeVGammaRays2024}) and fluxes (grey squares; \citealt{Peretti:2023crf}); MAGIC upper limits (brown diamonds; \citealt{MAGIC_4151}); the IceCube neutrino flux estimate (magenta region; \citealt{2025ApJ...981..131A}); and the COSI two-year continuum sensitivity (pink curve; \citealt{2024icrc.confE.745T}).\label{fig:proton_photon_neutrino_stat}}
\end{figure*}

\section{Results for NGC 4151} \label{sec:results}
We adopt $R_{\rm c} = 15\,R_{\rm g}$ \citep[with $R_{\rm g} \equiv GM_{\rm BH}/c^2$ the gravitational radius of black hole of mass $M_{\rm BH} = 10^{7.2}M_\odot$ in the case of NGC 4151; ][]{4151_mass}, consistent with the estimated upper bound on the corona size \citep{dasRevealingProductionMechanism2024}, $l_{\rm coh} = 4.0\,R_{\rm g}$, and $\xi_{p,\rm inj} = 0.1$.  We assume a disk luminosity $L_{\rm disk} = 2\times10^{43}$~erg\,s$^{-1}$ for NGC~4151, with a fraction $f_{\rm cor} = 0.2$ reprocessed by the corona, yielding an approximate X-ray luminosity $L_{\rm X} \sim 4\times10^{42}$~erg\,s$^{-1}$, in agreement with observations \citep{kumarNGC4151}. 

The energy distribution of secondary photons and electrons/positrons injected by hadronic and Bethe-Heitler interactions depends on the model parameters, in particular the ratios of advection or escape to acceleration timescales~\citep[][for details]{SLB26}. To bypass this ambiguity, we fix an acceleration rate $\nu_{\rm acc}$ such that, for $Z_\pm=0$, the secondary $\gamma$-ray flux saturates the existing bounds from Fermi-LAT at GeV energies. Such a coronal hadronic interpretation for the sub-GeV $\gamma$-ray flux observed in nearby Seyfert galaxies has been proposed by \cite{muraseSubGeVGammaRays2024} and \cite{Carpio_2026}. Our approach then provides a conservative upper bound to the contribution of hadronic electromagnetic cascades in the sub-GeV range.
When expressed in terms of a diffusion coefficient benchmarked on PIC simulations~\citep{comissoInterplayMagneticallyDominated2019a,wongFirstprinciplesDemonstrationDiffusiveadvective2020a, wongEnergyDiffusionAdvection2025}, our acceleration rate corresponds to a turbulent Alfvénic velocity $v_{\rm A} = 0.26\,c$, in the range of expected values (Sec.~\ref{sec:electrons}). The corresponding model also provides a reasonable account of the observed neutrino flux within its current uncertainties (see Fig.~\ref{fig:proton_photon_neutrino_stat}).

In the case of NGC~4151, we find that the $pp$ channel provides a dominant contribution to the secondary $\gamma-$ray flux, because of the comparatively low X-ray luminosity compared, e.g., to NGC~1068. Varying the pair loading factor $Z_\pm$, and hence the background proton density at a fixed optical depth, offers a way of probing the hadronic contribution to the sub-GeV $\gamma$-ray flux. Through complementary simulations, we have verified that this exploration of different $Z_\pm$ with other parameters held fixed covers well the possible spectral shapes of neutrino and $\gamma$-ray fluxes. 

Figure~\ref{fig:proton_photon_neutrino_stat}, which represents the main outcome of our study, presents the resulting multi-messenger spectrum for each of the electron prescriptions shown in Figure~\ref{fig:electron} and varying $Z_\pm$, overplotted on existing data for NGC 4151\footnote{The X-ray spectrum is corrected for absorption and reflection with \texttt{XSPEC} \citep{arnaud1996}. \textit{NuSTAR}, \textit{INTEGRAL}, and \textit{Swift}/BAT data are rescaled to the long-term \textit{Swift}/BAT average flux, for consistency with the two-year COSI sensitivity.}. Clearly, 
the non-thermal electron population leaves
a distinct imprint on the MeV spectrum. 

For a pure thermal primary electron distribution, the soft gamma-ray photon spectrum displays an exponential cutoff, then recovers at a low flux level through the hadronically-induced power-law tail extending down to the MeV-GeV band. However, the two-year COSI sensitivity threshold shown in Fig.~\ref{fig:proton_photon_neutrino_stat} forecloses a possible detection of this secondary hadronic emission, even for the maximal case saturating the sub-GeV $\gamma$-ray flux.  

The radiative output of the fiducial electron spectrum derived from the {\it ab initio} PIC simulation dominates the hadronic channel up to energies $\sim 10\,$MeV, and lies well within the reach of COSI at MeV energies. This conclusion remains unaltered when one considers the broader range of nonthermal primary electron spectra discussed in Sec.~\ref{sec:electrons}, provided their energy fraction verifies $\xi_{e,\rm nth}\gtrsim 0.05$. In the MeV to GeV range, 
the dominance of the leptonic contribution
gradually gives way to the hadronically-induced cascade.
The relative weight between the two components is set by the efficiency of hadronic interactions, controlled in this case by the value of $Z_\pm$, together with the energy fraction carried by nonthermal electrons. Our results show that a detailed measurement of the spectral shape in the range $0.01-1\,$GeV appears as the best discriminator between these populations. However, this will require, at least, 
a next generation of MeV instruments, such as AMEGO-X~\citep{2022JATIS...8d4003C} or e-ASTROGAM~\citep{2018SPIE10699E..2JT}. 
Note that the origin of the $0.1-100$\,GeV \textit{Fermi}-LAT emission toward NGC\,4151 remains debated, owing to possible contamination \citep{muraseSubGeVGammaRays2024,Peretti:2023crf}; Fig.~\ref{fig:proton_photon_neutrino_stat} includes both sets of measurements. Even if this emission originates from NGC\,4151, its physical origin remains disputed: it could arise from an ultra-fast outflow \citep{Peretti:2023crf,Neher:2026gct}, or from the corona. 

\begin{figure}
\centering
\resizebox{0.95\linewidth}{!}{\includegraphics[clip=true]{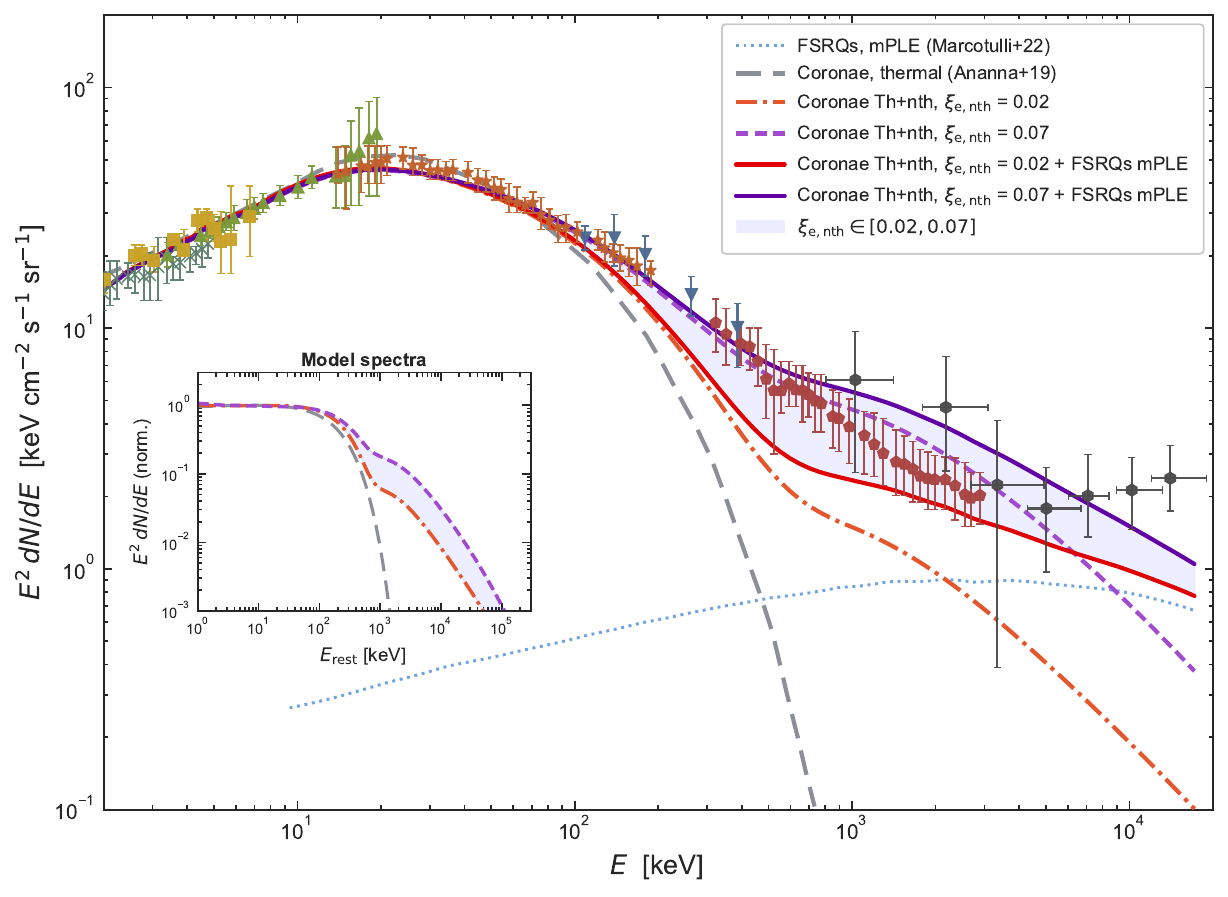}}
\caption{Predicted spectral shape of the diffuse cosmic X-ray background, for different
descriptions of the coronal nonthermal electron population. The grey dashed
line shows the thermal-only prediction of \cite{ananna2019}. The blue dotted
line is the blazar contribution from \cite{marcotulli2022} (mPLE model). Red
and purple dashed lines show our coronal prediction, following the
\cite{ananna2019} methodology but with a nonthermal electron spectrum, for
two values of $\xi_{\rm e,nth}$ (at fixed $p$). Solid red and purple lines show
the total (corona + blazars). The inset shows the model spectra underlying
each CXB prediction. Data taken from \citet{marcotulli2022}: \textit{Chandra}-COSMOS (yellow), \textit{ASCA} (grey), \textit{RXTE} (green), \textit{Swift}-BAT (orange), Nagoya balloon experiment (blue), \textit{SMM} (red), and COMPTEL (black).  \label{fig:cxb}}
\end{figure}

\section{The cosmic X-ray-MeV background} \label{sec:extension}

The sub-MeV excess of the cosmic X-ray background (CXB) is generally attributed to AGN, but has proven difficult to reproduce self-consistently: the X-ray component is typically attributed to thermal coronal electrons, while the higher-energy $\gamma$-ray emission is attributed to jetted flat spectrum radio quasars (FSRQs) \citep{marcotulli2022}, leaving an unexplained gap in between. The inclusion of a nonthermal coronal electron population can potentially bridge this gap and explain the MeV excess~\citep{2008ApJ...672L...5I,Inoue2018}.  

To place our predicted nonthermal coronal spectrum in the broader context of the CXB, we repeat the population-synthesis calculation of \citet{ananna2019}, using their best-fit X-ray luminosity function and our prescribed electron distributions. For torus reprocessing, we use \texttt{borus02} \citep{balokovic2018} instead of the original \texttt{BNTORUS} table \citep{brightman2011}, which is no longer publicly available. Using their intrinsic coronal continuum, described by a cut-off power law of (photon) index $\Gamma = 1.96$ and cutoff $E_{\gamma,\rm cut}=200$~keV, this procedure recovers their Figure~11 to within $\sim 10\%$. We then substitute their coronal continuum with a similar coronal spectrum computed with \texttt{Turb-AM3}, featuring a nonthermal electron tail ($p= 3.0$, $\gamma_{e, \rm cut}=10^3$) and two representative values of the nonthermal energy fraction, $\xi_{\rm e, nth} = 0.02$ and $0.07$. For the FSRQ contribution we adopt the mPLE model of \citet{marcotulli2022}, in which the characteristic luminosity evolves with redshift at a fixed space density. The shaded band of Fig.~\ref{fig:cxb} shows the range spanned by these two values of $\xi_{\rm e,nth}$, combined with this blazar contribution. The purely thermal model fails to reproduce the observed CXB excess at $\sim 0.1-1$~MeV, while $\xi_{\rm e, nth} \simeq 0.02$-$0.07$ brings the model into agreement with the data.

Our results thus demonstrate overall consistency, in that the same nonthermal electron population, informed from first-principles kinetic numerical simulations (Sec.~\ref{sec:electrons}), used to model the multi-messenger SED of NGC~4151, provide a satisfactory description of the $0.1-1$ MeV excess of the cosmic X-ray background. The best match is obtained for a nonthermal energy fraction $\xi_{e,\rm nth} \sim 0.05$, which lies at the detection threshold for the COSI mission. For this value, the hadronic electromagnetic cascades could dominate over the primary electron channel as early as a few MeV, in the most extreme case in which these hadronic cascades also saturate the existing measurements in the sub-GeV range. In the Appendix, we further apply our model to the X-ray binary Cygnus~X-1, and demonstrate that a similar nonthermal fraction provides a reasonable match to the observed MeV tail in the hard state.

\section{Conclusions} \label{sec:conclusions}
Motivated by recent {\it ab initio} kinetic simulations probing the energization physics of turbulent, magnetized AGN coronae, we used the \texttt{Turb-AM3} framework to self-consistently model the resulting X-ray-to-neutrino emission from NGC\,4151, and to compare it with the emission from the hadronically-induced electromagnetic cascade.

We find that the nonthermal electron population observed in these kinetic simulations leaves a distinctive imprint on the predicted MeV spectrum, erasing the exponential cutoff expected from a purely thermal corona and producing an extended emission component, well within the range of  sensitivity of COSI. For an extended family of primary nonthermal electron distributions characterized by smaller nonthermal fractions and larger cut-off energies, motivated by current limitations of these kinetic simulations, we find that the radiative output remains detectable by COSI provided the nonthermal energy fraction exceeds $\sim 5\%$. This nonthermal electron emission dominates the MeV band and substantially outweighs the hadronic component, which generically takes over at sub-GeV energies. The transition is set by the interplay between proton acceleration efficiency and the nonthermal electron fraction. Next-generation instruments are needed to better disentangle the leptonic and hadronic contributions in the $0.01-0.1\,$GeV range.

We have extended our analysis to the diffuse cosmic X-ray background and showed that the same nonthermal electron distribution informed by PIC simulations provides a natural explanation for the diffuse MeV excess, which lends further support to the present model. As briefly demonstrated in the accompanying Appendix, the same model also provides a satisfactory fit of the MeV tail in the stellar-mass black hole corona of Cygnus~X-1.

Together, these results show that nonthermal electrons, rather than being a secondary correction to the thermal Comptonizing population, can be an energetically significant and observationally distinct component of AGN coronae, whose signature is within reach of forthcoming MeV instruments. Upcoming COSI observations of NGC\,4151 will therefore provide a direct test of the turbulent particle acceleration physics at play in these coronae.

\begin{acknowledgments}
SLB  gratefully acknowledges helpful discussions with X. Rodrigues.
DG gratefully acknowledges helpful discussions with R.~Mushotzky.
This publication is based upon work from COST Action CA24149 -- Extreme Plasma Network for Advanced Discovery (EXPAND), supported by COST (European Cooperation in Science and Technology). SLB and ML work was supported by the French Agence Nationale de la Recherche, ANR, project ANR-25-CE31-3279.
DG~was supported by the Research Foundation--Flanders (FWO) Senior Postdoctoral Fellowship 12B1424N. \software{AM3 \citep{klingerAM$^3$OpenSourceTool2024}, XSPEC \citep{arnaud1996}, borus02 \citep{balokovic2018}, NumPy \citep{numpy2020}, SciPy \citep{scipy2020}, Matplotlib \citep{matplotlib2007}, Astropy \citep{ astropy2022}.}
\end{acknowledgments}

\appendix
\twocolumngrid
\section{Application to Cygnus~X-1}\label{app:cygx1}
\begin{figure}
\centering
\includegraphics[width=0.95\columnwidth]{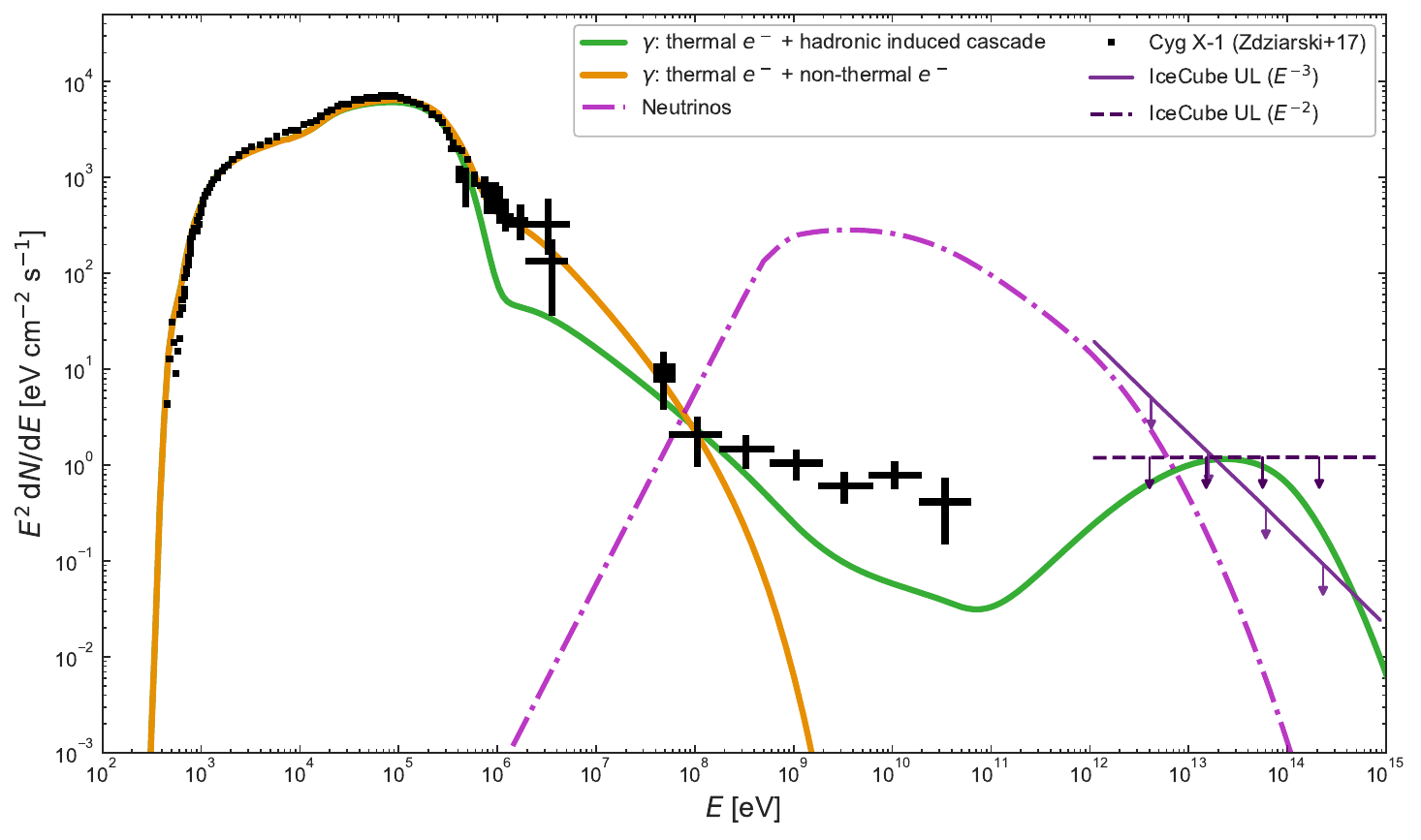}
\caption{Multi-messenger spectra predicted for the Cygnus~X-1 corona in the hard state. The orange curve shows the coronal emission from thermal and nonthermal electrons, without the hadronic secondary pair production. The green curve shows the coronal emission from thermal electrons alone, together with the hadronically-induced cascade. The dot-dashed purple line shows the corresponding neutrino production. Black points show the broadband spectral energy distribution of Cygnus~X-1 in the hard state, as measured by \textit{BeppoSAX}, \textit{INTEGRAL}, COMPTEL, and \textit{Fermi}-LAT \citep{Zdziarski2017}. The purple solid and grey dashed lines show the IceCube 90\% C.L. upper limits on the neutrino flux assuming $E^{-3}$ and $E^{-2}$ spectra respectively \citep{2022ApJ...930L..24A}. \label{fig:cyg_X1}}
\end{figure}

In its hard state, the $\gamma$-ray spectrum of Cygnus~X-1 exhibits two distinct high-energy tails, at $\sim 0.1$--$10$\,MeV and $0.1$--$10$\,GeV, whose origins remain debated~\citep[see][and references therein]{2025ApJ...993...54K}. The high polarization degree reported in the $0.4$--$2$\,MeV band~\citep{laurent2011} has been interpreted as evidence for a jet contribution~\citep{zdziarski2014,Zdziarski2017,kantzas2021}, although a coronal origin has also been advocated~\citep{mcconnell2002,2014SSRv..183...61P,romero2014,Fang24}, and hadronic channels may contribute as well~\citep[e.g.][]{2013MNRAS.434.2380M,Fang24,wei2025}. Leaving polarization aside, we focus here on the broadband spectrum of \citet{Zdziarski2017} and test whether the coronal model developed in Sec.~\ref{sec:electrons} can account for it.

To this end, we adopt a characteristic coronal size of \(R_c = 5\,R_g\) \citep{2024Galax..12...80J} and a black-hole mass of \(M_{\rm BH}=16\,M_\odot\) \citep{Orosz2011}. As for NGC~4151, we set the proton acceleration rate such that the hadronic $\gamma$-ray emission saturates the observed constraints at $\sim 0.1$\,GeV; using the same PIC-benchmarked scaling of the acceleration rate, this corresponds to an Alfv\'en velocity $v_{\rm A}\simeq 0.24\,c$. These parameters correspond to a magnetic field of $B\simeq9.6\times10^6$\,G \citep[compatible with the estimates of][]{2013MNRAS.430..209D} and an electron magnetization of \(\sigma_e\sim100\). We inject soft photons from the accretion disk and normalize their luminosity such that the resulting \(2\)--\(10\) keV X-ray luminosity is \(L_{2-10\,{\rm keV}}\simeq1.3\times10^{37}\) erg s\(^{-1}\) \citep[][]{Zdziarski2017}. For the adopted coronal size, this corresponds to a compactness of \(\ell\simeq43\), close to the value inferred by \citet{2015MNRAS.451.4375F}. We further adopt a turbulent coherence length \(l_{\rm coh}=R_c/3\) and a nonthermal proton injection fraction \(\xi_{\rm p,inj}=0.1\). We consider here two models for the electron population: one with thermal electrons only, $\Theta_e=0.2$; a second including a nonthermal component with $p=3$, $\xi_{\rm e, nth}=0.04$, and a high-energy cutoff $\gamma_{e,\rm cut}=10^3$. In this comparison, we switch off the hadronic processes in the nonthermal-electron model in order to isolate the leptonic and hadronic contributions.

The resulting spectra, corrected for absorption and reflection, are shown in Fig.~\ref{fig:cyg_X1}, together with the observed average hard-state spectrum of Cygnus~X-1. The thermal primary electron model including hadronic cascades reproduces the main X-ray emission but does not account for the observed MeV excess. By contrast, including a nonthermal electron component naturally produces an extended high-energy tail and provides a good description of the observed emission up to 0.1\,GeV. The required electron distribution is broadly compatible with those obtained in recent PIC simulations, and with the fraction needed to explain the cosmic X-ray background in the 0.1--1\,MeV band (Fig.~\ref{fig:cxb}). We also find that the hadronic cascade cannot account for the observed MeV tail without exceeding the constraints at $\sim0.1$\,GeV. Nevertheless, the hadronic secondary cascades can provide a substantial contribution to the GeV emission observed from Cygnus~X-1. Since this hadronic model saturates the $\gamma$-ray constraints at $\sim 0.1$\,GeV, the corresponding neutrino flux should be regarded as an upper bound on the coronal neutrino emission of Cygnus~X-1. In the TeV range, this maximal flux is comparable to current IceCube upper limits \citep{2022ApJ...930L..24A}. Overall our results are consistent with \citet{Fang24}, although their methodology is different.

\bibliography{Article_NGC4151_2026}
\bibliographystyle{aasjournalv7.1}
\end{document}